\documentclass[useAMS,usenatbib,a4paper]{mn2e}

\usepackage{psfig}
\usepackage{graphics}
\usepackage{graphicx}
\usepackage{url}
\title[Spectral Turn-Overs in Millisecond Pulsars]{Low-Frequency
  Spectral Turn-Overs in Millisecond Pulsars Studied from Imaging
  Observations} 

\author[M.~Kuniyoshi et.~al.]
{M.~Kuniyoshi,$^{1,2}$\thanks{E-mail:
    masaya.kuniyoshi@nao.ac.jp} J.~P.~W.~Verbiest,$^{3,1}$
  K.~J.~Lee,$^{4,1}$ B.~Adebahr$^1$, M.~Kramer$^1$ \newauthor and
  A.~Noutsos$^1$\\ 
  $^{1}$Max-Planck-Institut f\"ur Radioastronomie, Auf dem H\"ugel 69,
  53121 Bonn, Germany\\
  $^2$NAOJ Chile Observatory, National Astronomical Observatory of
  Japan, 2-21-1 Osawa, Mitaka, Tokyo 181-8588, Japan\\
  $^{3}$Fakult\"at f\"ur Physik, Universit\"at Bielefeld, Postfach
  100131, 33501 Bielefeld, Germany\\
  $^4$Kavli institute for astronomy and astrophysics, Peking
  University, Beijing 100871, P.R.~China\\}
\begin{document}

\date{Accepted. Received ; in original form }

\pagerange{\pageref{firstpage}--\pageref{lastpage}} \pubyear{2014}

\maketitle

\label{firstpage}

\begin{abstract}
  Measurements of pulsar flux densities are of great importance for
  understanding the pulsar emission mechanism and for predictions of
  pulsar survey yields and the pulsar population at large. Typically
  these flux densities are determined from phase-averaged ``pulse
  profiles'', but this method has limited applicability at low
  frequencies because the observed pulses can easily be spread out by
  interstellar effects like scattering or dispersion, leading to a
  non-pulsed continuum component that is necessarily ignored in this
  type of analysis. In particular for the class of the millisecond
  pulsars (MSPs) at frequencies below 200\,MHz, such interstellar
  effects can seriously compromise detectability and measured flux
  densities. In this paper we investigate MSP spectra based on a
  complementary approach, namely through investigation of archival
  continuum imaging data. Even though these images lose sensitivity to
  pulsars since the on-pulse emission is averaged with off-pulse
  noise, they are insensitive to effects from scattering and provide a
  reliable way to determine the flux density and spectral indices of
  MSPs based on both pulsed and unpulsed components. Using the 74\,MHz
  VLSSr as well as the 325\,MHz WENSS and 1.4\,GHz NVSS catalogues, we
  investigate the imaging flux densities of MSPs and evaluate the
  likelihood of spectral turn-overs in this population. We determine
  three new MSP spectral indices and identify six new MSPs with likely
  spectral turn-overs.
\end{abstract}


\section{Introduction}
It is well established that there are two fundamental classes of
pulsars: the normal (or ``young'') pulsars, with spin periods between
roughly a tenth of a second and a few seconds and spin-down
rates between $10^{-18}$ and $10^{-12}$s/s; and the class of the
millisecond pulsars (MSPs), with spin periods below about 30\,ms and
spin-down rates below $10^{-17}$s/s \citep{lgy+12}. According to
standard theory, this latter group is created through a ``recycling''
process where a normal pulsar accretes mass from a main-sequence
companion star through Roche-lobe overflow \citep{acrs82}. This
accretion process increases the spin of the pulsar but also affects a
variety of emission properties, as summarised by \citet{kxl+98}.

% Pulsars are roughly categorized into two groups consisting of MSPs
% with a period of $\sim$ millisecond and slow pulsars with $\sim$
% second.  It is considered that the former were originally slow
% puslars, but mass transfer from the binary companion turned the slow
% pulsars into MSPs (Alpar et al. 1982), which is in fact confirmed by
% the fact that MSPs are mostly seen in binary systems.  This means that
% the evolutionary history of MSPs is quite different from that of slow
% pulsars, whose difference emerges in their observed pulsar properties
% (e.g. pulsar periods, pulsar derivatives, characteristic age, magnetic
% field, etc.), but people think the emission mechanism having
% responsibility of radio emission is the same in both MSPs and slow
% pulsars, which is in fact supported by the comparison of both emission
% properties at high frequencies (Kramer et al. 1998, 1999).
% 
Studies of the spectral index of pulsars are of particular importance
to any attempts at understanding the pulsar emission mechanism; and to
clarify any differences between MSPs and their slower
counterparts. Even though only few MSP spectra were available,
previous work \citep{kxl+98,tbms98} found no significant difference
between the spectral index distributions of MSPs and young pulsars,
but \citet{kll+99} did find that MSP spectra tended to be somewhat
less steep. Interestingly, recent simulations by \citet{blv13}
indicate that spectral indices available for slow pulsars are biased
towards steeper spectra because most pulsars with determined spectral
indices were discovered in early, low-frequency surveys, which are
most sensitive to steep-spectrum objects. Arguably the same bias does
not exist for the MSPs, which were mostly discovered in 1.4-GHz
surveys and are less biased to steep-spectrum sources. Clearly, a
full comparison of the populations will require a significant increase
in the number of spectral indices measured for MSPs.

While spectral indices provide the simplest description of a spectrum,
some pulsar spectra display more complex features. Of specific
interest to our discussion are the spectral turn-overs that have been
observed in some pulsars at frequencies below 400\,MHz
\citep[e.g.][]{sie73,kms+78,ikms81} and at frequencies near 1\,GHz in a
smaller sample \citep{mkkw00a}.
%
% One of intriguing phenomena in slow pulsars can be a spectral
% turn-over at the low frequency roughly between 100(50) and 400 MHz
% (e.g. Siever (1973), Kuzmin et al. (1978), Izvekova et al. (1981),
% etc.), while there are two compelling rare slow pulsars B1838$-$04
% with a spectral turn-over discovered at as high a frequency as $\sim$
% 1 GHz (Maron et al. 2000).

Research into breaks in MSP spectra has been hampered by the
small amount of data available; and inconsistencies in this limited
data set. \citet{em85} investigated the spectrum of the first known
MSP (PSR~J1939+2134 or B1937+21) and detected an indication of a
possible turn-over in the spectrum below 75\,MHz. Subsequently,
\citet{ffb91} observed a sample of four MSPs, including
PSR~J1939+2134, at a variety of frequencies and found no significant
deviations from power-law spectra in any of these
pulsars. \citet{mabe96} published the spectrum of the brightest MSP
known, PSR~J0437$-$4715 and found clear evidence for a spectral break
around 200\,MHz. This pulsar remained the only MSP with a clear
spectral turn-over until \citet{kl01} published the largest sample so
far, analysing the spectra of 30 MSPs at frequencies down to
100\,MHz. Out of their large sample, only PSR~J1012+5307 was found to
have a turn-over in its spectrum. Most recently, observations of
MSP~J2145$-$0750 with the Long Wavelength Array in New Mexico
\citep{etc+13} have shown very clear evidence of a flattening of the
spectrum of this pulsar below a few hundred MHz \citep{drt+13},
suggesting that the non-detection of a break by \citet{kl01} was
caused by a lack of measurement points and sizeable uncertainties on
their measurements. This brings the total number of MSPs with
suggested spectral breaks to four, out of a total sample of 33
MSPs with spectra down to $\sim100$\,MHz.

As this historical overview shows, the spectra of MSPs have only
been investigated down to $\sim100$\,MHz on a small number of sources;
and when additional data were added to previously published spectra,
the results were contradicting as often as they were confirming. This
illustrates how lack of sensitivity; lack of observations; and the
lack of access to a truly wide frequency range have hampered this
field.

One effect that limits the accessible frequency range in which pulsar
flux density measurements can be made, lies in the way most of these
observations are conducted. Typical pulsar observations are performed
using the phase-folding technique, in which the incoming data are
averaged as a function of pulse phase, given the known rotational
period of the pulsar. This technique is normally advantageous because
it separates the pulsed signal from the off-pulse noise and thereby
gains significant sensitivity. However, when investigating flux
density measurements, this phase-folding technique implicitly assumes
that the off-pulse region (i.e. the phase-range where the pulsed
emission reaches a minimum) only contains noise and no emission
originating from the pulsar \citep{lk05}. This is not always true,
because if a pulsar has only a small offset between its rotation axis
and its magnetic axis, it is possible that the line of sight
permanently crosses the polar emission region. In this case even the
baseline level of the phase-folded profile would contain emission that
should be attributed to the pulsar, but which gets ignored as it is
unpulsed.

At low frequencies this situation gets worse, since the pulsar
emission is affected by a series of interstellar propagation effects
that tend to smear out the emission over pulse phase \citep[see][for a
review]{ric77}. Because dispersion can nowadays be corrected for
through coherent dedispersion, the most important of these effects is
scattering, which has approximately a $\nu^{-3.9}$ scaling
\citep{bcc+04} and is therefore particularly relevant at the low
frequencies where the spectral turn-overs are observed to occur.

An alternative approach to measuring spectra of MSPs is to determine
the pulsar's flux density from interferometric imaging. Since these
maps are not phase-resolved they do tend to have less sensitivity, but
in this way the \emph{total} flux density can be
determined\footnote{Note that phase-resolved measurements of flux
  density typically report the \emph{phase averaged} flux density
  after subtracting a noise floor determined at off-pulse phases. This
  means that phase-resolved flux densities are directly comparable to
  interferometric flux density measurements. The only exception is the
  case where some part of the pulsar's emission beam is continuously
  pointed towards us, as in this case only the pulsed fraction of the
  flux density is measured.} (including any potentially unpulsed
component) and this flux density measurement is entirely insensitive
to interstellar propagation delays like scattering. Moreover, several
all-sky imaging surveys across a variety of observing frequencies have
been released into the public domain, allowing spectrum investigations
on a large sample of sources without any further requirement for
observing time. Such imaging surveys were already used by
\citet{kcac98} who investigated pulsar flux densities and
scintillation based on the 1.4\,GHz NVSS (NRAO VLA Sky Survey) and the
325\,MHz WENSS (Westerbork Northern Sky Survey), but they did not
investigate these data for spectral breaks as they did not have access
to a truly low-frequency survey.
%
% As far as the flux measurements of MSPs at low frequency range,
% aperture synthesis radio survey images at low frequencies could shine.
% Unlike the phase-folding flux measurement for pulsed flux densities,
% synthesis images are generated by correlating signals from a
% collection of radio telescopes, producing total flux densities.
% Therefore, flux measurements by imaging have nothing to do with
% incorrect DM models and scattering effects, etc.  Moreover, there are
% fortunately already large-survey radio images available for the pulsar
% flux measurements including the 74MHz VLSSr, the 325 MHz WENSS, and
% 1.4 GHz NVSS (see the next section for details).
% 
% In fact, large-scale flux measurements of pulsars\footnote{Most of
%   them are slow pulsars.} were performed by Kaplan et al. (1998),
% showing a comparison of the NVSS 1.4 GHz flux densities and PSRCAT 1.4
% GHz flux densities $S_{1400 MHz}$ (Tayloar, Manchester, and Lyne 1993)
% yields reasonable agreement, generating spectral indices together
% together with the 325 MHz WENSS or PSRCAT $S_{400MHz}$.
% 

Since the analysis by \citet{kcac98}, the known sample of pulsars has
more than doubled, and with the recent publication of the improved
74\,MHz VLSSr \citep[VLA Low-Frequency Sky Survey Redux,][]{lchk12}, the
sensitivity of a similar investigation is significantly enhanced at
the lower frequencies, providing significantly improved sensitivity to
spectral breaks over an unprecedentedly large number of sources.

In this paper, we present the results from an investigation into
spectral turn-overs in MSPs, based on the 74\,MHz VLSSr,
complemented with the 325\,MHz WENSS and the 1.4\,GHz NVSS. These
three surveys are described in Section~\ref{sec:explanation_images},
along with the selected source sample. Section~\ref{sec:results} lists
our detections and the obtained spectra; and in Section~\ref{sec:conc}
we summarise our findings.
%
% Since then, a lot of pulsars including many MSPs have been discovered,
% and the VLSS was performed in 2007, and then its image quality has
% enhanced with improvement of algorithms in data reduction, and renewed
% as the 74 MHz VLA Low-Frequency Sky Survey Redux (VLSSr) in 2012,
% allowing us to have opportunities to search for the spectral turn-over
% in MSPs, which is often observed in most of the slow pulsars.  Section
% 2 describes the identification method and summarize each survey radio
% image we used for this study.  Section 3 describes the results with
% investigation of the spectral turn-over in MSPs.  In section 4, we
% discuss the results.

\section{Identification with Aperture Synthesis Radio Survey
  Images}\label{sec:explanation_images}
In order to investigate the spectra of MSPs, we investigated the data
from the three imaging surveys discussed below, for detections of
MSPs. In this context we defined MSPs as being pulsars with spin
periods below 30\,ms and a time-derivative of the spin period
$\dot{P} < 10^{-17}$s/s\footnote{We also considered pulsars with
  as-yet undetermined spindown $\dot{P}$, since these undetermined
  values are likely to be small.}. We excluded pulsars in globular
clusters to avoid obvious issues with confusion; and restricted our
analysis to pulsars at declinations above $-40\degr$, as this is the
lowest declination limit amongst the three surveys used. A list of
candidate sources and their positions was taken from the ATNF pulsar
catalogue\footnote{Available at
  \url{http://www.atnf.csiro.au/research/pulsar/psrcat}.}
\citep{mhth05} and based on those positions the images from the three
surveys were inspected for detections. In some cases the positions
from the catalogue are referred to epochs that are up to a decade or
more after the data for the surveys were taken. This could cause the
pulsar position to have changed due to the high velocity of
pulsars. However, based on proper motion measurements of the MSPs in
our sample (where available) or the average transverse velocity of the
pulsar population \citep[which is less than 100\,km/s,][]{hllk05},
none of the MSPs in our sample was likely to have moved as much
as a resolution element in any of the surveys used.

Identification of our candidate sources was done through queries of
the available source catalogues%of the 74\,MHz VLSSr and the 1.4\,GHz
%NVSS surveys
\footnote{The catalogue browsers for these surveys can be found at
  \url{http://www.cv.nrao.edu/vlss/VLSSlist.shtml},
  %http://heasarc.gsfc.nasa.gov/W3Browse/radio-catalog/wenss.html 
  \url{http://www.astron.nl/wow/testcode.php?survey=1} and
  \url{http://ww.cv.nrao.edu/nvss/NVSSlist.shtml} respectively.} and
through visual inspection of the survey images, where a 3-$\sigma$
detection threshold coincident with the catalogue position of the
pulsar was required to claim a detection. For the sources below
$\sim 5\sigma$ significance, which were not in the online catalogues,
the position and flux density were subsequently determined through
Gaussian fitting to the images. For all of our detections the offset
between the fitted position and the most recent published position was
well within the resolution of the images.

As mentioned, three imaging surveys were used. These were chosen based
on their observing frequency, sensitivity and fraction of sky
covered. The surveys used were:
\begin{description}
\item[VLSSr:] The original Very Large Array (VLA) Low-Frequency Sky
  Survey (VLSS) and its catalogue were released in 2007
  \citep{clc+07}. It covers 95$\%$ %of 3$\pi$ sterradians
  of the sky north of $-$30\degr declination at a frequency of 74\,MHz
  with a resolution of approximately 80\arcsec\ and an RMS sensitivity
  of about 100\,mJy/beam. The minimum integration time on each field
  is 75 minutes, consisting of three shorter observations, each
  approximately 25 minutes, separated in time by at least one hour.
  Following improvements in various data reduction algorithms
  (particularly related to bright-source peeling, radio frequency
  interference removal and ionospheric calibration) a renewed release
  of the survey and images was made in 2012 \citep{lchk12}.
\item[WENSS:] The Westerbork Northern Sky Survey (WENSS), was
  performed with the Westerbork Synthesis Radio Telescope (WSRT),
  which consists of 14 parabolic antennas, between 1991 and 1996 at a
  frequency of 325\,MHz, covering the entire sky north of declination
  $+28.5$\degr with a limiting 1-$\sigma$ flux density of
  approximately 3.6\,mJy. This survey has an imaging resolution of
  54$\arcsec\times 54\arcsec/ \sin\left(\delta\right)$ with $\delta$
  the declination \citep{rtd+97}. 
  % and a positional accuracy of approximately 1.5\arcsec
\item[NVSS:] The NRAO VLA Sky Survey (NVSS) was performed with the VLA
  at a frequency of 1.4\,GHz with the compact D and DnC configurations
  during 1993 and 1996, with some additional observations carried out
  in 1997. The NVSS is a continuum survey covering the entire sky
  north of $-$40\degr declination with a FWHM angular resolution of
  45$\arcsec$ and 1-$\sigma$ RMS flux density fluctuations of
  $\approx$0.45\,mJy/beam \citep{ccg+98}. 
  % , which is nearly uniform sensitivity for all the images.
\end{description}

\section{Results}\label{sec:results}
According to the ATNF catalogue\footnote{Version 1.50 of the ATNF
  catalogue, used on 25 July 2014.}, 130 sources satisfied the
selection criteria laid out in Section~\ref{sec:explanation_images}.
Out of these, we detected ten pulsars in the VLSSr data. A further
six pulsars were expected to be detectable based on an extrapolation
of the available higher-frequency data, but were not detected,
providing strong evidence for a spectral break. The remaining 113
sources were not detected, but did not indicate a spectral
turn-over. In some cases this was the case because their spectral
index (as derived from higher-frequency data) is flat enough to make a
non-detection at 74\,MHz expected; while for other pulsars there are
no flux density measurements at multiple frequencies available, making
it impossible to derive a spectral index and therefore yielding no
expected flux density at 74\,MHz. The third and largest group (47 out
of 113) were not detected because they lay outside the survey area,
because the VLSSr data for them were corrupted, or because their
positions are not known precisely enough to allow a confident
identification.
%
%either because they were not expected to be detectable (based on the
%available higher-frequency data and spectral index) or because they
%had insufficient higher-frequency data available to allow an
%extrapolation to lower frequencies. 47 of these 113 non-detected
%pulsars were not detected in the VLSSr data because they lay outside
%the survey are or the data were corrupted, or the position was not
%known precisely enough yet to allow a confident detection.

In the following, we define the spectral index $\alpha$ as in 
\begin{equation}\label{eq:spindx}
S\left(\nu\right) \propto \nu^{\alpha},
\end{equation}
with $S\left(\nu\right)$ the flux density at frequency $\nu$.  Since
fitting non-linear functions in the presence of uncertainties is
complicated and does not reliably quantify the often asymmetric
uncertainties, we applied a Monte Carlo approach to the determination
of our spectral indices. Practically this means we drew random
realisations from the flux density measurements at all frequencies
(using the published flux density values and interpreting the
measurement uncertainties as Gaussian and symmetric).
%using their uncertainties and most likely values and
%modelling the uncertainties as Gaussian), and then used an unweighted
A power-law function (Equation~\ref{eq:spindx}) was then fitted to
these simulated measurements through a linearised unweighted fit to
determine realistic initial values which were used as the basis of a
non-linear, weighted fit.
%and finally fit to determine the best linear regression to these
%realisations in a logarithmic space. 
The distribution of the slopes of these final fits (shown in figures
\ref{fig.four-three}, \ref{fig.three_spectral_index} and
\ref{fig.turnover}), then provided the most likely spectral index and
its uncertainties.

In the cases where a spectral turn-over was required, this was always
clear without requiring a spectral fit, i.e.\ no borderline cases were
found. In these cases a separate spectral slope was fitted to the
lowest- and highest-frequency measurements, up to a break frequency
that was visually identified.

%
% 128 MSPs in the Galactic plane were derived from the ATNF pulsar
% catalog, and we successfully found MSP candidates in the 74 MHz VLSSr
% images/catalogs and identified them with the MSPs by taking advantage
% of the information of the pulsar positions found in the VLSSr and
% wheather the candidate flux densities are on a simple power-law
% trendline Eq.(1) inferred based on a collection of published flux
% measurements \footnote{ we assumed a typical uncertainty of 30$\%$ for
%   the data where the authors did not mention the error uncertainty
%   level.  }.
% \begin{equation}
% S_{f}=A\cdot e^{B} 
% \end{equation}
% $S_{f}$ is the flux density at frequency $f$, A is the coefficient and
% B is the spectral index.  For the gaining of the power-law trendline,
% we first generated fluxes consisting of the published flux
% measurements and a gaussion noise corresponding to their flux
% uncertainty level.  As it is arduous to perform a non-linear fitting,
% we took a logarithm in the equation (1) to get a linear equation and
% obtained both A and B values by perforiming a least-squares fitting to
% data.  Then we used them as the initial values for the non-linear
% fitting for the equation (1), gaining a coefficient A and spectral
% index B.  We performed this procesure 100,000 times for creating a
% probability density function, determinging the trustful power-law
% trendline.
 
For the MSPs that were not detected in the VLSSr data, we used the
same Monte Carlo approach on any available high-frequency data but
extended the approach to provide a distribution of expected flux
densities at an observing frequency of 74\,MHz, assuming the power-law
flux density scaling of Equation~\ref{eq:spindx}. This distribution
was then compared to the 3-$\sigma$ upper limit on the flux density
from the VLSSr images to evaluate the likelihood of a spectral
turn-over in these pulsars.
% 
% As for the MSPs with no VLSSr candidates, whose flux densities at 74
% MHz are however deduced to be above 3$\sigma$ noise level of the 74
% MHz VLSSr image from the spectral index based on the data published
% for higher frequencies, we inspected them and succeeded in producing
% some MSPs with high possibility of exhibiting the spectral turn-over.
% The results were divided into several categories as the following
% subsections below.  As far as WENSS and NVSS candidates, we used those
% mainly as a supplement to confirm if there is no discrepancy in the
% determined spectral indices.

In the following sections, we discuss our detections and the
non-detections that may indicate spectral turn-overs. Specifically, in
Section~\ref{sec:7det} we discuss six detected MSPs that are
consistent with previously published data; in Section~\ref{sec:4det}
we present three detected MSPs that have their spectral index
determined for the first time; and in Section~\ref{sec:7nondet} we
discuss the six MSPs whose non-detections indicate likely spectral
turn-overs.

%\subsection{Seven MSPs on the spectral trendline based on the
%  published data}\label{sec:7det}
\subsection{Six MSPs Detected and Consistent with Previous
  Publications}\label{sec:7det}
\begin{figure*}
  % \epsscale{1.8}
  %\psfig{angle=0.0,height=20cm,figure=four13-2.eps}
  %\psfig{angle=0.0,width=18cm,figure=figure1_combined.eps}
  %\psfig{angle=0.0,width=18cm,figure=Figure1.eps}
  \includegraphics[width=18cm,angle=0.0,trim= 10 15 0 0]{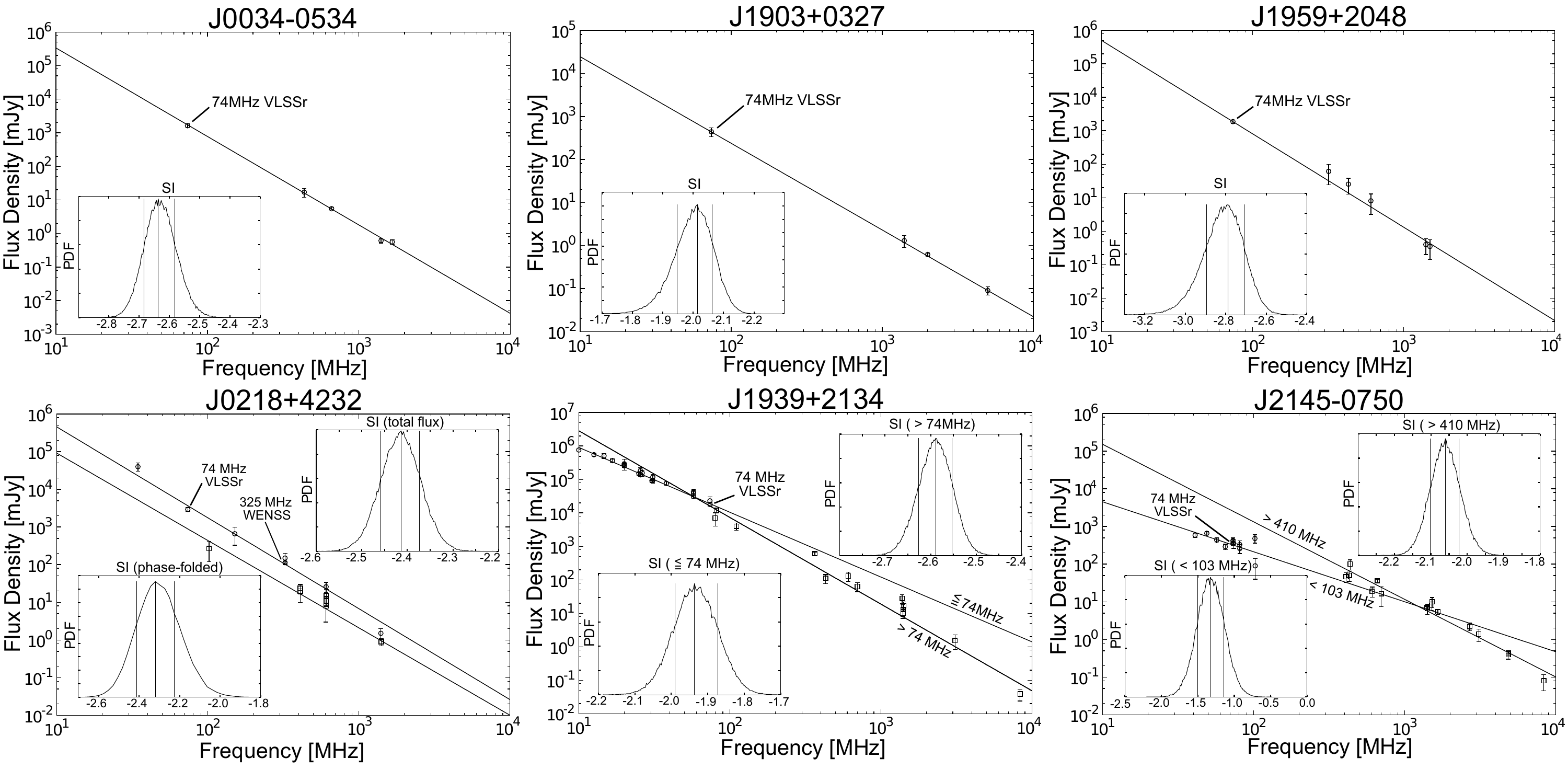}
  % \plotone{four13-2.eps}
  \caption{Spectra for the six detected pulsars with existing
    spectral information. Top row: spectra for three MSPs
    with no sign of a spectral break. Bottom row: spectra for the
    three MSPs with more complex spectra. The spectra for
    PSRs~J1939+2134 and J2145$-$0750 show spectral turn-overs; and the
    spectrum for 
    PSR~J0218+4232 is consistent but at a higher level in imaging data 
    than in phase-folded data as a consequence of the unpulsed
    emission (see Section~\ref{ssec:0218} for details). The insets
    show the probability density functions for the spectral index
    distribution for each pulsar; in these insets the most likely
    spectral index and its 1-$\sigma$ confidence interval are
    indicated by vertical bars.
    For the three pulsars on the bottom row, two different spectra
    were fitted. These two fits were based on the flux density measurements
    shown with dashed symbols; and flux density measurements shown in full
    symbols. In practice, for PSR~J0218+4232, this means that dashed
    measurements are from phase-folded work, while the full markers
    indicate flux densities derived from imaging data.
    % Left column: 4 MSPs extended the spectra toward lower frequencies,
    % showing no spectral turn-over. Right column: 3 MSPs on the trend
    % lines inferred by the published data. In SI (spectral index)
    % panels, the red line shows the most likely spectral indiecs, bule
    % and green lines are 1$\sigma$ uncertainty level.
  }
    \label{fig.four-three}
\end{figure*}

In total, we detected 10 pulsars in the VLSSr. Four of these have no
previous spectral index published and will therefore be discussed
separately, in Section~\ref{sec:4det}. The remaining six sources are
all consistent with previously published spectral indices derived from
higher-frequency data. Of these six, only PSR~J0218+4232 was detected
in the WENSS (the non-detections being primarily caused by the limited
sky coverage of this survey) and the low luminosity of MSPs at higher
frequencies made that only PSR~J2145$-$0750 was detected in the
NVSS. The list of the detected pulsars and all the flux density values
used in our determination of spectral indices, is given in
Table~\ref{tab:fluxes}, the spectra of these pulsars are shown in
Figure~\ref{fig.four-three} and the derived spectral indices are given
in Table~\ref{tab:alpha}. For a few sources a further discussion of
our results is given below.
\begin{table*}
  \begin{center}
    \caption{Summary of flux density measurements used. This table
      contains the flux densities used for our determination of
      spectral indices (summarised in Table~\ref{tab:alpha}),
      combining measurements from the VLSSr, WENSS and NVSS surveys
      with those already available from literature. In cases where the
      measurement uncertainty was not clear from the publication, we
      assumed a 30\% uncertainty. These measurements are indicated by
      an asterisk.
    }
    \label{tab:fluxes}
    \begin{tabular}{l|rrl}
      \hline
      Pulsar name  & frequency & flux density  & Reference and Notes \\
                   & (MHz)     & (mJy)         &                     \\
      \hline
      J0034$-$0534 &   74      & $1620\pm210$  & This work.    \\
                   &  103      &  $250\pm120$  & \citet{kl01},
                                                 significantly lower
                                                 flux density, not used in fit\\
                   &  436      &   $17\pm5$    & \citet{tbms98}\\
                   &  660      &  $5.5\pm0.5$  & \citet{tbms98}\\
                   & 1400      & $0.61\pm0.09$ & \citet{tbms98}\\
                   & 1660      & $0.56\pm0.09$ & \citet{tbms98}\\

      J0218+4232 &  102      & $270\pm150$   & \citet{kl01}\\
      (pulsed)   &  410      &  $24\pm7$*    & \citet{stc99}\\
                 &  410      &  $20\pm10$    & \citet{nbf+95}\\
                 &  606      &   $8\pm5$     & \citet{nbf+95}\\
                 &  610      &  $15\pm7$     & \citet{kxl+98}\\
                 &  610      &  $11\pm3$*    & \citet{stc99}\\
                 & 1410      &   $0.9\pm0.2$ & \citet{nbf+95}\\
                 & 1410      &   $0.9\pm0.2$ & \citet{kxl+98}\\

      J0218+4232 &   34.5    & $(4\pm1)\times10^4$& \citet{du90}  \\
      (total)    &   74      & $2940\pm360$  & This work.\\
                 &  151      & $660\pm330$   & \citet{nbf+95}\\
                 &  325      &  $113\pm4$    & This work.\\
                 &  325      & $150\pm50$    & \citet{nbf+95}\\
                 &  608      & $26\pm7.8$    & \citet{nbf+95}\\
                 & 1400      & $1.5\pm0.5$   & \citet{nbf+95}\\

      J1810+1744 &   74      & $1090\pm150$   & This work.\\
                 &  350      & $20\pm4$      & \citet{hrm+11}\\

      J1816+4510 &   74      & $690\pm110$    & This work.\\
%                 &  350      & $1.5\pm0.45$*   & STOVALL ET AL. 2014\\
%                 &  820      & $0.28\pm0.084$* & STOVALL ET AL. 2014\\
%                 &  325      &  $11\pm4$     & This work.\\

      J1843$-$1113 &   74      & $820\pm194$   & This work.\\
                 & 1400      & $0.10\pm0.02$  & \citet{hfs+04}\\

      J1903+0327 &   74      & $440\pm100$   & This work.\\
                 & 1400      & $1.3\pm0.4$   & \citet{crl+08ltd}\\
                 & 2000      & $0.62\pm0.05$ & \citet{crl+08ltd}\\
                 & 5000      & $0.09\pm0.02$ & \citet{crl+08ltd}\\

      J1939+2134 & 74        & $17940\pm2150$  & This work.\\
      (B1937+21) & various   & various       & See \citet{em85} for a complete listing\\
                 & 700       & $63\pm19$     & \citet{mhb+13ltd}\\
                 & 1400      & $13\pm5$      & \citet{mhb+13ltd}\\
                 & 2695      & $2.0\pm 0.4$  & \citet{kll+99}\\
                 & 3000      & $1.6\pm0.7$   & \citet{mhb+13ltd}\\
                 & 4850      & $1.0\pm0.2$   & \citet{kll+99}\\
                 & 8350      & $0.039\pm0.015$ & \citet{kkmj12} \\

      J1959+2048 &   74      & $1880\pm240$   & This work.\\
      (B1957+20) &  318      & $60\pm40$     & \citet{fbb+90} (retrieved from their Figure~2)\\
                 &  430      & $26\pm13$     & \citet{fbb+90} (retrieved from their Figure~2)\\
                 &  606      & $8\pm5$       & \citet{fbb+90} (retrieved from their Figure~2)\\
                 & 1490      & $0.4\pm0.2$   & \citet{fbb+90} (retrieved from their Figure~2)\\
                 & 1400      & $0.4\pm0.2$   & \citet{kxl+98}\\
      \hline
    \end{tabular}
  \end{center}
\end{table*}
\begin{table*}
  \begin{center}
    \contcaption{}
    \begin{tabular}{l|rrl}
      \hline
      Pulsar name  & frequency & flux density  & Reference and Notes \\
                   & (MHz)     & (mJy)         &                     \\
      \hline
    J2145$-$0750 & various   & various       & See \citet{drt+13}.\\
                 & 74        & $354\pm91$   & This work.\\
                 & 102       & $480\pm120$   & \citet{kl01}\\
                 & 102.5     & $90\pm50$     & \citet{mms00}\\
                 & 410       & $46\pm14$*    & \citet{stc99}\\
                 & 430       & $50\pm15$*    & \citet{bhl+94}\\
                 & 436       & $100\pm30$    & \citet{tbms98}\\
                 & 610       & $19\pm6$*     & \citet{stc99}\\
                 & 660       & $36\pm4$      & \citet{tbms98}\\
                 & 700       & $16\pm9$      & \citet{mhb+13ltd}\\
                 & 1400      & $7\pm0.9$     & \citet{tbms98}\\
                 %& 1400      & $9\pm13$      & \citet{mhb+13ltd}\\
                 & 1400      & $2.6\pm0.4$   & This work;
                 significantly lower flux density, not used in fit\\
                 & 1414      & $6.6\pm2.0$*  & \citet{stc99}\\
                 & 1510      & $8\pm2$       & \citet{kxl+98}\\
                 & 1520      & $10\pm3$*     & \citet{bhl+94}\\
                 & 1660      & $5.4\pm0.8$   & \citet{tbms98}\\
                 & 2695      & $2.2\pm0.5$   & \citet{kll+99}\\
                 & 3100      & $1.4\pm0.5$   & \citet{mhb+13ltd}\\
                 & 4850      & $0.4\pm0.1$   & \citet{kll+99}\\
                 & 4850      & $0.44\pm0.03$ & \citet{kkwj97}\\
                 & 8350      & $0.080\pm0.037$ & \citet{kkmj12} \\

      J2215+5135 & 74        & $800\pm120$   & This work.\\
%                 & 325       &   $6.6\pm3.6$ & This work.\\
                 & 350       &   $5\pm1.5$*  & \citet{hrm+11}\\
      \hline
    \end{tabular}
  \end{center}
\end{table*}
\begin{table}
  \begin{center}
    \caption{Summary of spectral index measurements. This table
      contains the spectral indices derived from the measurements in
      Table~\ref{tab:fluxes}.}
    \label{tab:alpha}
    \begin{tabular}{lcrc}
      \hline
      Pulsar       & Spectral              & Turn-over     & $\alpha$ \\
      name         & index $\alpha$        & frequency $\nu_{\rm TO}$ & below $\nu_{\rm TO}$ \\
      \hline
      J0034$-$0534 & $-2.64\pm0.05$        & --            & -- \\
      J0218+4232   & $-2.41\pm0.04$        & --            & -- \\
      J1810+1744   & $-2.57^{+0.14}_{-0.16}$ & --            & -- \\
%      J1816+4510   & $-3.76\pm0.16$        & --            & -- \\
      J1843$-$1113 & $-3.08\pm0.10$ & --            & -- \\
      J1903+0327   & $-2.01^{+0.07}_{-0.05}$ & --            & -- \\
      J1939+2134   & $-2.59\pm0.04$        & $\sim74$\,MHz & $-1.94^{+0.06}_{-0.05}$ \\
      J1959+2048   & $-2.79^{+0.08}_{-0.11}$ & -- & -- \\
      J2145$-$0750 & $-2.60\pm0.04$        & $\sim400$\,MHz & $-1.33^{+0.18}_{-0.17}$\\
      J2215+5135   & $-3.22^{+0.15}_{-0.23}$ & --            & -- \\
      \hline
    \end{tabular}
  \end{center}
\end{table}

\subsubsection{PSR J0034$-$0534}\label{ssec:0034}
The spectral index of this pulsar has been measured twice before, by
\citet{tbms98} who derived a spectral index of $\alpha = -2.6\pm1.0$
and subsequently by \citet{kl01} who found a slightly shallower
spectrum with $\alpha = -2.3\pm 0.3$. Our VLSSr detection lowers the
lowest frequency in the spectrum from 111\,MHz to 74\,MHz and is in
perfect agreement with the earlier \citet{tbms98} measurements, but in
significant disagreement with the flux density value of
\citet{kl01}. This disagreement indicates the \citet{kl01} flux
density for this pulsar may have been underestimated. Excluding the
\citet{kl01} value from our analysis, we derive a spectral index of
$\alpha = -2.64\pm0.05$. We note furthermore that the folded
pulse profile for this pulsar is very wide at these low
frequencies\footnote{See, e.g. the European Pulsar Network Database at
  \url{http://www.jb.man.ac.uk/research/pulsar/Resources/epn/browser.html}.},
making it possible that some component of unpulsed, scattered flux
density was unaccounted for in the \citet{kl01} measurement.

\subsubsection{PSR J0218+4232}\label{ssec:0218}
\citet{nbf+95} discovered PSR~J0218+4232 and reported a significant
fraction of its radio emission was not pulsed, implying it is an
aligned rotator and causing an offset between its total flux density
(as derived from imaging) and its pulsed flux density (as derived from
phase-folded observations). \citet{stc99} subsequently determined the
magnetic inclination angle to be consistent with 0\degr, based on a
fit of the rotating vector model \citep[RVM, see][]{rc69a} to
observations at 410 and 610\,MHz, thereby confirming the
classification as an aligned rotator.

We detected PSR~J0218+4232 in the VLSSr data with a flux density of
$2.9\pm0.4$\,Jy. In the WENSS, we detected the pulsar at 113\,mJy. We
did not find the pulsar in the NVSS, which may be explained by
scintillation \citep{ric77} since the survey's 1.35\,mJy 3-$\sigma$
sensitivity is close to the 1.5\,mJy published flux density at
1.4\,GHz.

%  The VLSSr candidate
% with the flux of 2.94 Jy ($\sigma$ $\approx$ 100 mJy) was detected
% 6.35$\arcsec$ away from the ATNF timing position (Du et al. 2014).
% The WENSS counterpart with approximately 114 mJy was also found at the
% position $\Delta$RA=$-0.22651^{s}$ and $\Delta$DEC=$-2.8143^{\arcsec}$
% away from the ATNF position \footnote{WENSS(B1950):
%   RA=$02^{h}14^{m}58.566^{s}$,
%   DEC=$42^{\circ}18^{\arcmin}30.86^{\arcsec}$, ATNF(B1950):
%   RA=$02^{h}14^{m}58.33949^{\arcsec}$,
%   DEC=$42^{\circ}18^{\arcmin}28.0457^{\arcsec}$}.  Apart from those

Combining our own measurements and those from \citet{nbf+95} and
\citet{stc99} for both imaging and phase-folded cases, we find no
evidence for a spectral turn-over and determine a spectral index
$\alpha = -2.41\pm0.04$ for the total flux density, which is
consistent with the spectral index for the phase-folded flux density:
$\alpha = -2.32\pm0.09$, showing that the unpulsed flux
density evolves consistently with the pulsed flux density (see
Figure~\ref{fig.four-three}).

\subsubsection{PSR J1939+2134}\label{ssec:1939}
This pulsar was originally known as 4C21.53 and was detected in many
aperture synthesis images before being uncovered as the first MSP
\citep[then known as PSR~B1937+21,][]{bkh+82}. Consequently, its
spectrum has been studied extensively, as can be seen from our summary
in Figure~\ref{fig.four-three}, having flux density measurements from
4.8\,GHz all the way down to 10\,MHz. An initial controversy about the
presence of a spectral turn-over could be explained by the very slight
nature of the turn-over and the very low frequencies at which it
occurs.

We detected PSR~J1939+2134 in the VLSSr data with a flux density of
$17.94\pm2.15$\,Jy. This measurement is consistent with previously
published flux density measurements and confirms the slight spectral
turn-over first indicated by \citet{em85}. The WENSS did not cover
this position and a confident detection in the NVSS was made
impossible by a coincident extended source of emission slightly north
of the pulsar position.

% 4C21.53 was observed many times by aperture synthesis interferometers
% and known as an enigma source with the spectral index approximately -2
% for many years.  Backer et al. (1982) detected millisecond
% periodicity, leading to the discovery of the first millisecond pulsar,
% B1937+21 (J1939+2134).  As shown in the spectral index chart, there
% have been enough observations to determine the spectral index from 10
% to 1400 MHz (see Fig.\ref{fig.four-three}).  
% 
% We also found the VLSSr
% candidate with the flux of 17.94 Jy ($\sigma$ $\approx$ 90 mJy) very
% close to the ATNF timing position \cite{cognard95} with the angular
% separations 5.33$\arcsec$.  The flux density is consistent with what
% is inferred by the other publsihed data showing this is considered to
% have a spectral break at around 74 MHz (Erickson et al. 1985).  The
% NVSS candidate was not detected because an extended source a bit north
% from the pulsar position dominates this area.  The WENSS did not cover
% this area. 
%
% POSSIBLY ADD IN LOFAR COMPARISON HERE
% 
%  Kondratiev et al. (2014) detected J1939+2134, but however
% its flux density was far below the expected flux density derived from
% other data.  They concluded it was brought on by scattering.  With a
% clear detection with the VLSSr, we believe that image flux
% measurements become more important when studying PSR J1939+2134 at
% lower frequencies.

\subsubsection{PSR J1959+2048}\label{psr1959}
PSR~J1959+2048 (also known as PSR~B1957+20) is the first known
eclipsing pulsar, discovered at 430\,MHz by \citet{fst88}. The
eclipses occur when the pulsar passes behind its degenerate,
0.1-$M_{\odot}$ companion, which is being ablated by the pulsar
wind. At the discovery frequency, the pulsar is eclipsed $\sim$10\% of
the time, but this eclipse fraction scales strongly with wavelength
\citep{sbl+01}.

We detected PSR~J1959+2048 in the VLSSr data with a flux density of
$1.88\pm0.24$\,Jy. This area was not covered in the WENSS and we did
not detect the pulsar in the NVSS either, possibly due to
eclipses. Our measurement extends the spectrum further down,
indicating the absence of spectral breaks.
%
%  and
% immediate monitored by Fruchter et al. (1988), which revealed that
% J1959+2048 disappears in eclipses by the companion for nearly 10$\%$
% orbit: this pulsar is categorized into Black Widows having very low
% mass companions (less than 0.1 solar mass) which is degenerate
% (Mallory 2012).  
% 
%We found a plausible VLSSr candidate with a flux
%density value of as high as 1.88 Jy ($\sigma$ $\approx$ 90 mJy),
%9.79$\arcsec$ away from the ATNF position (Arzoumanian et al. 1994)
%and derived the spectral index $-$2.79(8) with the flux of this VLSSr
%candidate and the published data at higher frequencies from 318 MHz to
%1.49 GHz (Fruchter et al. 1990 and Kramer et al. 1998).  We did not
%find the 1.4 GHZ NVSS candidate for this pulsar, which can be caused
%by eclipses of the companion star.  The WENSS did not cover this area.
%
% POSSIBLY ADD IN LOFAR COMPARISON HERE
% 
% Hassall et al. (in preparation) conclude J1959+2048 showed a spectral
% turn-over with the LOFAR HBA at the frequency range between 100 to 200
% MHz, but however the 74 MHz VLSSr flux is utterly inconsistent with
% the one at 74 MHz extrapolated by the LOFAR HBA data.  As known from
% Fig.\ref{fig.sample_image} and Fig.\ref{fig.four-three}, there is an
% obviousness of the VLSSr candidate being J1959+2048.  We concluded
% that scattering effects might have caused mis-flux measurements in the
% LOFAR HBA.  For the flux measurements of J1959+2048 at this low
% frequency range, we conclude image flux measurements can be more
% trustworthy.

\subsubsection{PSR J2145$-$0750}\label{ssec:2145}
\citet{drt+13} recently observed PSR~J2145$-$0750 with the Long
Wavelength Array (LWA) in the frequency range 41 to 81\,MHz. Combining
their measurement with archival higher-frequency data, they fitted the
pulsar's spectrum to the following, somewhat unconventional,
functional form:
% leading
%to their conclusion that there is a spectral break at neary 730 MHz by
%fitting data with a power-law including spectral curvature of the
%form: 
\begin{equation}
S(\nu)=S_{0}\left(\frac{\nu}{\nu_{r}}\right)^{m_{1}+m_{2}\log(\nu/\nu_{r})}
\end{equation}
with $m_1$ the spectral index at the lowest frequencies, $m_{2}$ the
spectral curvature and $\nu_{r}$ the rollover frequency, which was
relatively arbitrarily chosen by \citet{drt+13} to be about 730\,MHz.

We detected PSR~J2145$-$0750 in the VLSSr image with a flux density of
$354\pm91$\,mJy. This flux density measurement is consistent with the
measurement by \citet{drt+13}, who determined a flux density of
$390\pm65\pm35\,$mJy at 73\,MHz, where the given uncertainties are for
measurement noise and systematic uncertainties, respectively. We also
detected this pulsar in the NVSS image, having a flux density of
2.6\,mJy, which is significantly less than the $7.0\pm0.9$\,mJy of
\citet{tbms98}. This discrepancy could be explained by scintillation,
as indicated by the long-term monitoring of \citet{mhb+13ltd}, who found
the flux density of PSR~J2145$-$0750 at 1.4\,GHz to have an RMS
scatter of no less than 12.5\,mJy. The pulsar's position was not
included in the WENSS.

Because no flux density measurements are published for this pulsar
between $\sim$100\,MHz and $\sim$400\,MHz, the exact break frequency
is ill-determined \citep[if there is indeed a particular break
frequency rather than the continuous turn-over advocated
by][]{drt+13}. Nevertheless, it is clear that at frequencies below
100\,MHz the spectrum is flatter than above 400\,MHz. Based on our own
and the archival data listed above, we determine spectral indices of
$-1.33_{-0.17}^{+0.18}$ and $-2.60\pm0.04$ respectively.
%
% We found the 74 MHz VLSSr
% candidate with a flux density of 356 mJy ($\sigma$ $\approx$ 80 mJy),
% 17.64$\arcsec$ away from the ATNF position (Verbiest et al. 2009),
% where we determined the flux and position with a Gaussian fitting,
% turning out that the flux density of the candidate is cosistent with
% that of LWA at 73 MHz. 
%  We also found the NVSS candidate within the
% error circle and determined the flux density of 2.4 mJy by a Gaussian
% fit, which could be lessened by sincillation as Toscano et al. (1998)
% reported this MSP has 7$\pm0.9$ mJy and Manchester et al. (2013)
% 8.9$\pm$12.5 mJy.  The WENSS did not cover this area.  
% 
% We roughly
% separated the published flux measurements including the VLSSr flux
% density into two areas below and above 103 MHz.  We are not sure where
% the roll-over starts but clearly confirmed there is a differnce in
% spectral indices between these two (see Fig.\ref{fig.four-three}).  We
% support the belief that there is a break in this pulsar In fact,
% Kondratiev et al. (in preparation) and Hassall et al. (in preparation)
% detected J2145$-$0750 with the LOFAR HBA and LBA band.  They also
% conclude this shows a spectral turn-over.

\subsection{Three New Spectral Indices}\label{sec:4det}
\begin{figure*}
  % \epsscale{1.0}
  %\psfig{angle=0.0,height=20cm,figure=three-new-spectral-index-v8.eps}
  %\psfig{angle=0.0,width=18cm,figure=4New.eps}
  %\includegraphics[width=18cm,angle=0.0]{Figure2.eps}
  \includegraphics[width=18cm,angle=0.0]{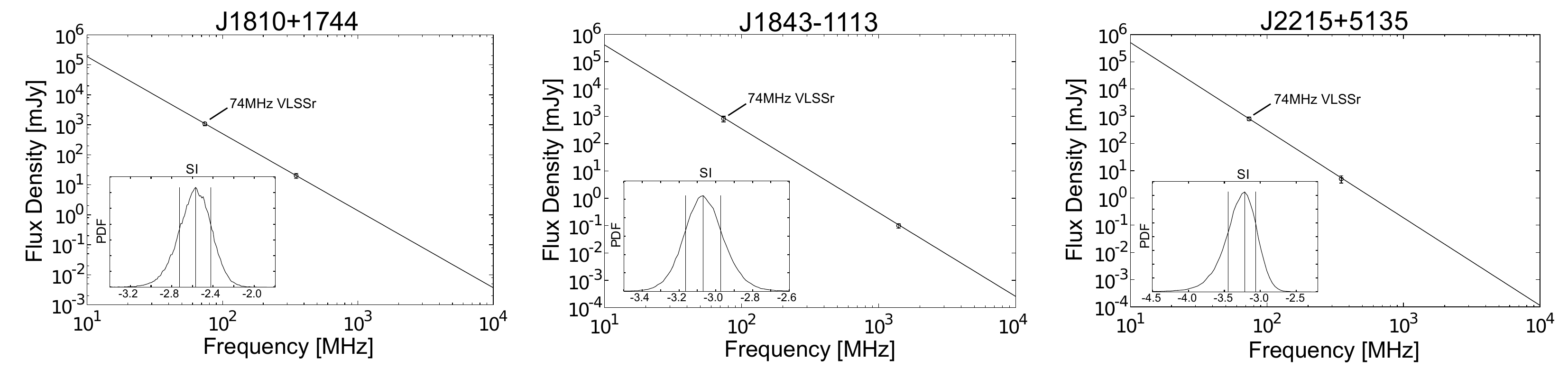}
  %\psfig{angle=0.0,width=18cm,figure=Figure2.ps}
  % \plotone{three-new-spectral-index-v8.eps}
  \caption{Three new MSP spectra obtained from our own imaging flux densities
    and published data. The inset panels show the likelihood distribution
    of the spectral indices, with the vertical lines indicating the
    most likely value and its 1-$\sigma$ interval.
    % panels, the red
    % line shows the most likely spectral indiecs, bule and green lines
    % are 1$\sigma$ uncertainty level. 
    % Each histogram used for the spectral index is shown in the bottom left corner of a panel, where
    % the red vertical and horizontal blue lines indicate the inferred spectral index, and   
    % 1 $\sigma$ confidence level.
  }
  \label{fig.three_spectral_index}
\end{figure*}

Four pulsars we detected in the VLSSr only had previous flux
density measurements at a single frequency (or none at all), implying
we were able to derive three spectral indices for the
first time. These three spectra are shown in
Figure~\ref{fig.three_spectral_index} and the flux density
measurements and spectral information are summarised in
tables~\ref{tab:fluxes} and \ref{tab:alpha}
respectively. Interestingly, two of the four pulsars in this class,
PSRs~J1810+1744 and J2215+5135 were found in a follow-up survey of
unidentified $\gamma$-ray sources in the Fermi all-sky map
\citep{hrm+11} and three out of four (PSRs~J1810+1744, J1816+4510 and
J2215+5135) inhabit eclipsing binary systems
\citep{hrm+11,ksr+12ltd}. We cannot investigate possible spectral
turn-overs for these pulsars, as we have less than three flux density
measurements per pulsar. For PSR~J1816+4510 specifically, we do not
determine a spectral index either, as the only flux density
measurements currently available are derived from uncalibrated survey
observations and are therefore unreliable \citep[see][and their
Figure~6 in particular]{slr+14ltd}\footnote{If we were to use the
  \citet{slr+14ltd} flux density values, an extreme spectral index of
  $-3.76$ would result, well in excess of the steepest pulsar spectral
  index currently published \citep[-3.5 for
  PSR~J0711+0931,][]{lzb+00}.}.

\begin{figure}
  \centering 
  \includegraphics[width=8.5cm,angle=0.0]{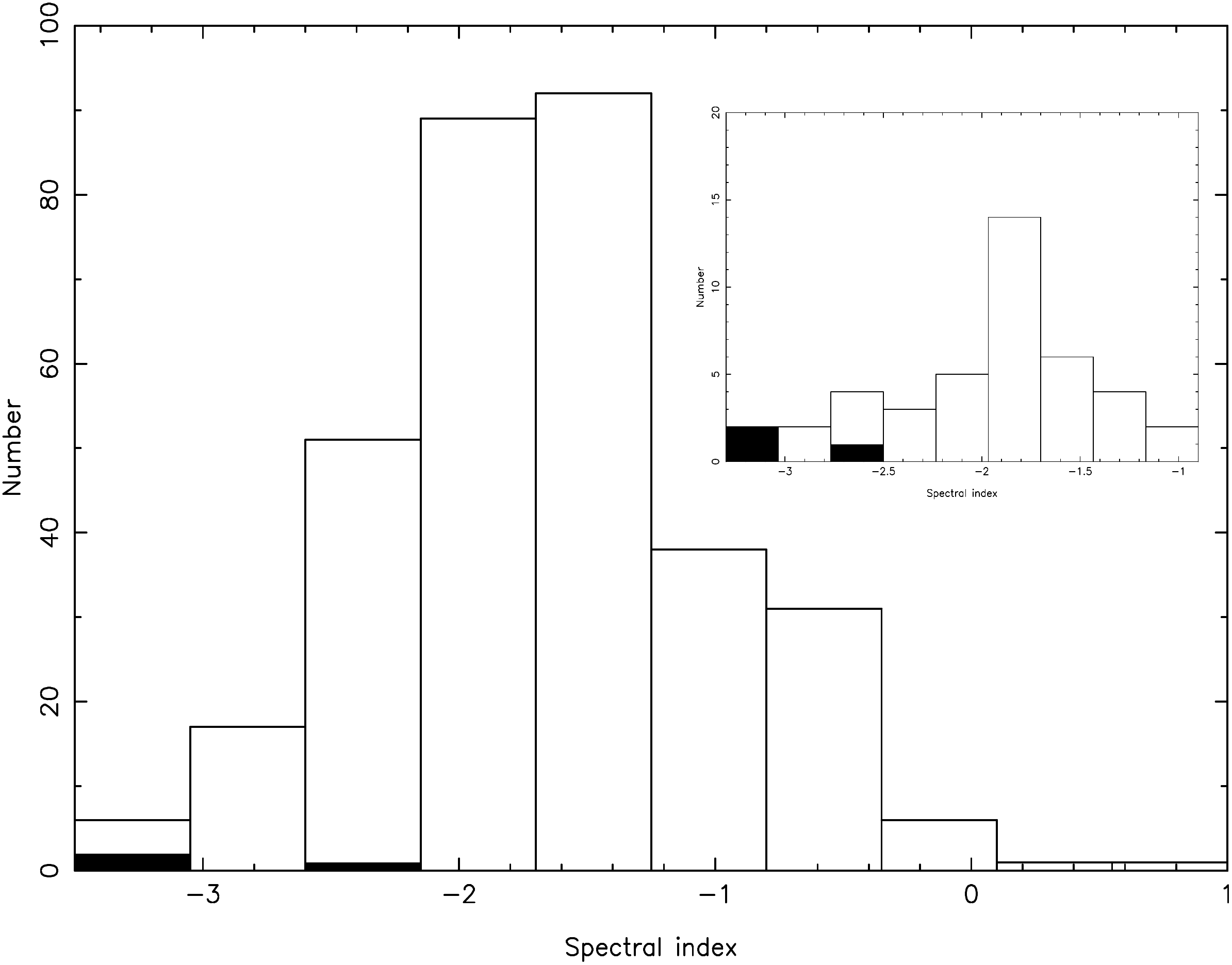}
  \caption{Histogram of all known pulsar spectral indices. The main
    figure shows all known spectral indices of pulsars, the inset
    shows the same plot for the class of the millisecond pulsars
    only. Our three new measurements are shown in black.}
  \label{fig:Hist}
\end{figure}

Finally, given their detections at low frequencies, these pulsars are
by design biased towards steep spectral indices. The measured spectral
indices are between $\alpha = -2.57$ and $\alpha = -3.22$, confirming
that these pulsars constitute a particularly steep-spectrum part of
the pulsar population: of the 328 spectral indices available in the
ATNF pulsar catalogue, only 9\% have equally steep spectra as these
three MSPs (see Figure~\ref{fig:Hist}).

\subsection{Six MSPs with a High Probability of Spectral
  Turn-Over}\label{sec:7nondet} 
\begin{figure*}
  %% \epsscale{1.8}
  %\psfig{angle=0.0,height=20cm,figure=turnover.eps}
  %\psfig{angle=0.0,width=18cm,figure=7Turnover.eps}
  \includegraphics[width=18cm,angle=0.0]{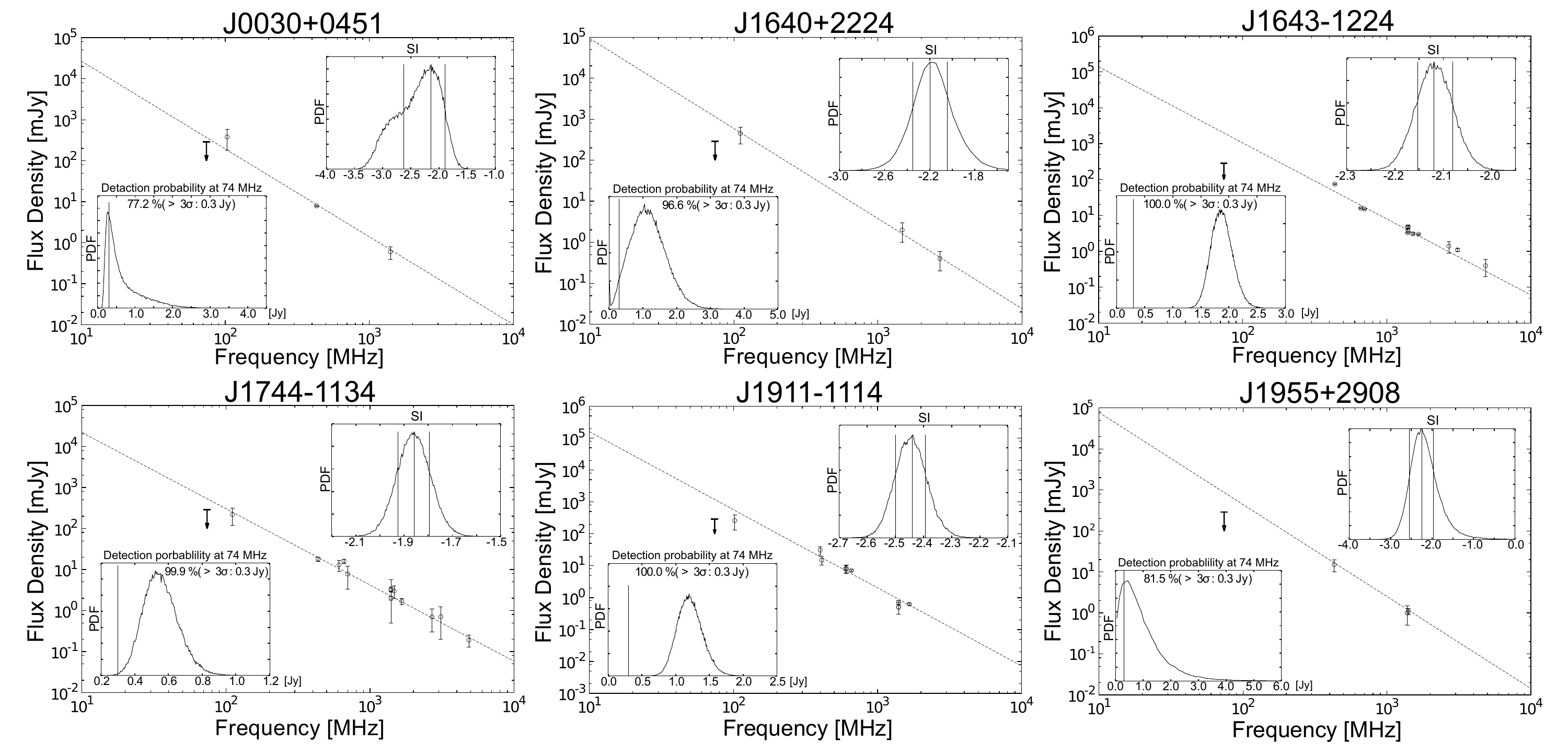}
  %\psfig{angle=0.0,width=18cm,figure=Figure3.ps}
  %% \plotone{turnover.eps}
  \caption{Spectra for six MSPs with a high probability of having a
    spectral turn-over. The arrows indicate the 3-$\sigma$ upper
    limits in the 74\,MHz VLSSr maps. The insets in the top-right of
    each sub-figure show the probability distribution of the spectral
    index based on the higher-frequency data referenced in
    Table~\ref{table_6breaks}; while the inset in the lower-left of
    each sub-figure shows the probability density function of the
    74\,MHz flux density based on this distribution of spectral
    indices, along with the detection probability at 74\,MHz assuming
    the spectrum does not have a spectral break. The vertical lines in
    the top-right plot show the most likely spectral index value along
    with its 1-$\sigma$ uncertainty interval; the vertical line in the
    lower-left plot shows the 3-$\sigma$ detection limit of the VLSSr
    (approximately 300\,mJy).
    %% \caption{Six MSPs with a high possibility of the spectral
    %% turn-over.
    %% The arrow in each spectral index chat indicates 3$\sigma$
    %% uncertainty level in the 74MHz VLSSr.
    %% The red line in the detection probability at 74 MHz indicates
    %% 300 mJy ($\approx $3$\sigma$ noise level
    %% in the VLSSr).
    %% In SI (spectral index) panels, the red line shows the most
    %% likely spectral indiecs, bule and green lines are 1$\sigma$
    %% uncertainty level.
    %% % Each histogram used for the spectral index is shown in the
    %% % bottom left corner of a panel, where
    %% % the red vertical and horizontal blue lines indicate the
    %% % inferred spectral index, and
    %% % 1 $\sigma$ confidence level.
  }
  \label{fig.turnover}
\end{figure*}

Based on fluxes at higher frequencies (either available from
literature, or derived from the NVSS or WENSS data), a spectral index
can be derived and used to predict the flux at 74\,MHz. For six
undetected pulsars this predicted flux lay above the 3-$\sigma$
detection limit of the VLSSr. This may in some cases be explained by
scintillation or because the higher-frequency flux density estimates
could be unreliable \citep{lbb+13ltd}.
%
%For seven pulsars in our sample, the spectral index derived from
%previously published flux densities (and in two cases a new NVSS or
%WENSS flux density), indicates an expected flux density at 74\,MHz
%that lies above the 3-$\sigma$ detection limit of the VLSSr. In some
%cases such pulsars may be missed because of scintillation or because
%the higher-frequency flux density measurements may be unreliable
%\citep{lbb+13}. 
With the exception of those caveats, a non-detection can also imply a
spectral turn-over. For these unexpectedly non-detected pulsars, we
determined the probability of a spectral turn-over using the Monte
Carlo approach outlined at the very start of this
section. Table~\ref{table_6breaks} provides the list of these pulsars,
along with the spectral indices derived from the published flux
densities and the likelihood of spectral turn-over following from our
analysis. For PSR~J1744$-$1134 we also used our NVSS detection at
$3.5\pm0.5$\,mJy. % and PSR~J1949+3106 was detected in the WENSS with a
%flux density of $23\pm6$\,mJy. 
The spectra are shown in Figure~\ref{fig.turnover}.

\begin{table*}
  \begin{center}
    \caption{Summary table for seven MSPs with a high probability of
      having a spectral break. Given are the pulsar name, the spectral
      index derived from previously published flux densities, the
      probability of turn-over and the literature references for the
      flux density measurements. Note these probabilities do not
      account for scintillation or underestimated uncertainties in the
      published measurements. References are as follows: (1)
      \citet{kl01}; (2) \citet{kxl+98}; (3) \citet{kll+99}; (4)
      \citet{lnl+95}; (5) \citet{tbms98}; (6) \citet{mhb+13ltd}; (7)
      \citet{stc99}; (8) \citet{llb+96}; (9) \citet{dfc+12ltd}; (10)
      \citet{bbf83}; (11) \citet{lzb+00}; (12) \citet{kcac98}.}
    \label{table_6breaks}
    \begin{tabular}{lcrl}
      \hline
      Pulsar       & Spectral              & Turn-over  & References\\
      name         & index $\alpha$        & likelihood &           \\
      \hline
      J0030+0451   & $-2.14^{+0.25}_{-0.52}$ &  77\% & 1, 11\\
      J1640+2224   & $-2.18_{-0.12}^{+0.12}$ &  97\% & 1, 2, 3\\
      J1643$-$1224 & $-2.11_{-0.04}^{+0.03}$ & 100\% & 2, 3, 4, 5, 6, 12\\
      J1744$-$1134 & $-1.85_{-0.08}^{+0.06}$ & 100\% & This work and
      1, 2, 3, 5, 6, 7\\
      J1911$-$1114 & $-2.45_{-0.05}^{+0.06}$ & 100\% & 1, 2, 5, 7, 8\\
      J1949+3106   & $-3.18_{-0.17}^{+0.24}$ & 100\% & This work and 9\\
      J1955+2908   & $-2.29_{-0.25}^{+0.34}$ &  82\% & 2 and 10\\
      \hline
    \end{tabular}
  \end{center}
\end{table*}
% (1) Kuzmin and Losovsky 2001, -> kl01
% (2) Kramer et al. 1998,       -> kxl+98
% (3) Kramer et al. 1999,       -> kll+99
% (4) Lorimer et al. 1995,      -> lnl+95
% (5) Toscano et al. 1998,      -> tbms98
% (6) Manchester et al. 2013,   -> mhb+13
% (7) The 1.4 GHz NVSS,
% (8) Stairs et al. 1999,       -> stc99
% (9) Lorimer et al. 1996,      -> llb+96
% (10) The 325 MHz WENSS, 
% (11) Deneva et al. 2012,      -> dfc+12
% (12) Boriakoff et al. 1984    -> bbf83

\section{Conclusions}\label{sec:conc}
There are two fundamentally different ways to determine pulsar flux
densities. These are through observations averaged modulo the
rotational period of the pulsar (``phase-folded'' measurements); and
through interferometric imaging observations. The former type has
higher sensitivity because of the separation of pulsed signal and
unpulsed noise, but the latter approach is unaffected by scattering in
the interstellar medium (which especially affects low-frequency data)
and does provide a correct measurement of the total pulsar flux
density even in the case of aligned rotators, where only a part of the
emission is pulsed.
% 
% There are two types of pulsar flux measurements comprising the
% traditional phase-folding for the pulsed flux and the imaging for the
% total flux.  These fluxes are same when pulsars are not aligned
% rotators, but however the phase-folding measurement is easily suffered
% by interstellar scattering and incorrect DM model values, etc.
% Especially at low frequencies, the phase-folding flux measurement
% could not work at worst.

In this paper, we have used archival imaging data to determine the
spectra of MSPs. We identified pulsars in the 1.4\,GHz NVSS, the
325\,MHz WENSS and, most importantly, the 74\,MHz VLSSr. The VLSSr is
of particular importance because it extends the investigated spectra
to the lower frequencies, where phase-folded observations are often
hampered by scattering effects and where investigation of potential
spectral turn-overs becomes possible.

We only obtained a few detections in the NVSS, mostly
because of the limited sensitivity of this survey compared to the low
flux densities of MSPs at 1.4\,GHz. In the WENSS our detection rate was
lowered by the survey's limited sky coverage. The VLSSr did
yield ten new pulsar flux density measurements; and for six sources a
non-detection in the VLSSr provided significant evidence for
spectral turn-overs at lower frequencies. Such evidence is hard to
obtain from phase-folded observations since those cannot distinguish
between pulsars that are too weak to be detected and pulsars that are
bright but heavily scattered.

In total, we extended the spectra of 16 pulsars to frequencies below
100\,MHz and found evidence for turn-overs in eight of these. Five of
the eight pulsars with spectral turn-overs were previously reported to
not have a spectral turn-over, based on higher-frequency data of
\citet{kl01}, suggesting in these cases the turn-over occurs near or
just below 100\,MHz. Following our work, a total of 39 MSPs have had
their spectra investigated at frequencies down to or below 100\,MHz;
and 10 out of those show evidence for turn-overs -- a much higher
fraction than previously thought.

The largest analysis of spectra for young pulsars to date, was
published by \citet{mkkw00}. They found there were two separate
populations of pulsars with spectral breaks. About 10\% of the
population displayed spectral steepening at frequencies around or
above 1.4\,GHz, whereas a far smaller number of pulsars (on the order
of a percent) showed spectral turn-overs at frequencies around
100\,MHz. The frequencies of the spectral turn-overs found in our
sample are comparable to the latter, low-frequency turn-over
sub-population. The fraction of the population displaying turn-overs
(about one in four), however, is similar to the fraction of young
pulsars showing high-frequency turn-overs ($\sim10\%$). This lends
some credence to the postulate that MSP spectra are fundamentally
identical to the spectra of young pulsars, albeit shifted to lower
frequencies, as first proposed by \citep{kll+99}.

Future complements to our work will be provided by a host of new SKA
pathfinder arrays that operate at the lowest frequencies continuously
observable from Earth. All three of these pathfinders (LOFAR, LWA,
MWA) are now capable of performing phase-resolved pulsar observations
\citep{sha+11ltd,drt+13,bot+14ltd} and will therefore carry spectral
investigations to lower frequencies, even for fainter pulsars that are
undetectable in most imaging surveys. Furthermore, all-sky surveys by
these three telescopes, combined with phase-resolved observations may
shed some light on the prevalence and strength of interstellar
scattering in the sample of MSPs that are undetectable in
phase-resolved observations.
\section*{Acknowledgments}
The authors wish to thank Kuo Liu and the referee, R.~N.~Manchester
for helpful comments on the manuscript, which improved the clarify and
correctness. K.~J.~Lee gratefully acknowledges support from the
National Basic Research Program of China, 973 Program, 2015CB857101
and NSFC 11373011.  This research has made use of the VLSS-Redux
Postage Stamp Server
%\footnote{http://www.cv.nrao.edu/vlss/VLSSpostage.shtml} 
and the VLSS-Redux Source Catalog
Browser\footnote{\url{http://www.cv.nrao.edu/vlss/}}; the NVSS Postage Stamp
Server
%\footnote{http://www.astron.nl/wow/testcode.php?survey=1}, 
and the NVSS Source catalog
browser\footnote{\url{http://www.cv.nrao.edu/nvss/}}. The National Radio
Astronomy Observatory (NRAO) is operated by Associated Universities,
Inc. and is a Facility of the National Science Foundation. This
research has also made use of the
WENSS\footnote{\url{http://www.astron.nl/wow/testcode.php?survey=1}} survey
data and we are grateful to the WENSS team consisting of Ger de Bruyn,
Yuan Tang, Roeland Rengelink, George Miley, Huub Rottgering, Malcolm
Bremer, Martin Bremer, Wim Brouw, Ernst Raimond and David Fullagar.
The WENSS project was a collaboration between the Netherlands
Foundation for Research in Astronomy (NWO) and the Leiden Observatory.

\bibliographystyle{mn2e}
\bibliography{journals,psrrefs,modrefs,crossrefs}

%% \clearpage
%% \begin{figure*}
%% \epsscale{1.8}
%% \plotone{sample_image.eps}
%% \caption{Sample Fits images with cross-mark MSP candidate positions.
%% \label{fig.sample_image}}
%% \end{figure*}
%% 
%% \begin{deluxetable}{lclcc}
%% \tabletypesize{\scriptsize}
%% \tablecaption{Three new spectral indices\label{table3}}
%% \tablewidth{0pt}
%% \tablehead{
%% \colhead{PULSAR} & DM &\colhead{VLSSr} & \colhead{Spectral} & \colhead{Reference} \\
%% \colhead{NAME} & (cm$^{-3}$pc) & \colhead{(mJy)} & \colhead{Index}
%% }
%% 
%% \startdata
%% J1810+1744\tablenotemark{B} & 39.7 & 1090(90)\tablenotemark{c} & $-2.58_{-0.12}^{+0.15}$\tablenotemark{*} & 1,2 \\
%% J1816+4510\tablenotemark{R} & 38.8 & \,\,\,690(90)\tablenotemark{c} & $-2.80_{-0.08}^{+0.10}$\tablenotemark{*} & 1,3 \\
%% J1843$-$1113 & 60.0 & 550(150)\tablenotemark{g} & $-2.92_{-0.11}^{+0.10}$\tablenotemark{*} & 1,4 \\
%% \enddata
%% \tablecomments{
%% References. (1) 74 MHz VLSSr catalog, (2) Hessels et al. 2011, (3) 325 MHz WENSS image, (4) Hobbs et al.2004
%% }
%% \tablenotetext{B}{Black Widows; there is a possiblity of an eclipse.}
%% \tablenotetext{R}{RedBacks; there is a possiblity of temporary no emission at radio.}
%% \tablenotetext{c}{The value was derived by the catalog.} 
%% \tablenotetext{g}{The value was derived by Gaussian fitting.} 
%% \tablenotetext{*}{The first-time spectral index.}
%% \end{deluxetable}

\label{lastpage}

\end{document}